\documentclass[prb, 11pt,superscriptaddress,floatfix,nofootinbib,notitlepage,longbibliography]{revtex4-1}
\usepackage[utf8]{inputenc}
\usepackage[T1]{fontenc}
\usepackage{lmodern}
\usepackage{graphicx}
\usepackage[colorlinks=true,citecolor=blue]{hyperref}
\usepackage{amsmath}
\usepackage{amssymb}
\usepackage{soul}
\usepackage{gensymb}
\usepackage{amsfonts,amssymb,amsbsy}
\usepackage{siunitx}
\usepackage{float}
\usepackage{physics}
\usepackage{booktabs}
\usepackage[percent]{overpic}

\usepackage{xcolor}
\usepackage{todonotes}

\newcommand{\RNum}[1]{\uppercase\expandafter{\romannumeral #1\relax}}

\usepackage{lineno}

\begin{document}

\title{Universality of Type-II Multiferroicity in Monolayer Nickel Dihalides}

\author{Aleš Cahlík}
\thanks{These authors contributed equally.}
\affiliation{Aalto University, Department of Applied Physics, 00076 Aalto, Finland}

\author{Antti Karjasilta}
\thanks{These authors contributed equally.}
\affiliation{Aalto University, Department of Applied Physics, 00076 Aalto, Finland}

\author{Anshika Mishra}
\affiliation{Aalto University, Department of Applied Physics, 00076 Aalto, Finland}

\author{Robert Drost}
\affiliation{Aalto University, Department of Applied Physics, 00076 Aalto, Finland}

\author{Mohammad Amini}
\affiliation{Aalto University, Department of Applied Physics, 00076 Aalto, Finland}

\author{Javaria Arshad}
\affiliation{Aalto University, Department of Applied Physics, 00076 Aalto, Finland}

\author{Büşra Arslan}
\affiliation{Aalto University, Department of Applied Physics, 00076 Aalto, Finland}

\author{Peter Liljeroth}
\affiliation{Aalto University, Department of Applied Physics, 00076 Aalto, Finland}

\begin{abstract}

The recent discovery of type-II multiferroicity in monolayer NiI$_2$ indicated a new pathway for intrinsic magnetoelectric coupling in the two-dimensional limit. However, determining whether this phenomenon is a unique anomaly or a general, chemically tunable property of the material class remains unresolved. Here, we demonstrate the universality of type-II multiferroicity in the transition metal dihalides by visualizing the ferroelectric order in monolayer NiBr$_2$. Using scanning tunneling microscopy (STM), we resolve atomic-scale ferroelectric domains and confirm their magnetoelectric origin through reciprocal manipulation experiments: reorienting magnetic order via electric fields and suppressing the electric polarization with external magnetic fields. Furthermore, we find that the multiferroic state in NiBr$_2$ is energetically less robust than in its iodide counterpart, consistent with modified superexchange interactions and the reduced spin-orbit coupling. Our results establish the transition metal dihalides as a versatile platform where the stability of magnetoelectric phases can be engineered through chemical substitution.

\end{abstract}

\date{\today}

\maketitle

\section*{Introduction}

A defining feature of two-dimensional (2D) van der Waals materials is their chemical tunability \cite{ajayan_two-dimensional_2016}. Just as the electronic bandgap in transition metal dichalcogenides can be engineered by substituting the chalcogen atom (e.g., S, Se, Te), the physical properties of magnetic monolayers can be precisely tuned through ligand substitution \cite{burch_magnetism_2018, Li2019, Wang2020}. This is exemplified in the transition metal dihalide (TMDH) family, where replacing the halogen ion (Cl, Br, I) systematically modifies the magnetic exchange interactions and spin-orbit coupling (SOC) strength without altering the crystal symmetry \cite{haavisto_topological_2022, fumega2022, riedl2022, Xu2024, Deng2025}. This chemical simplicity offers a control knob to navigate the magnetic phase diagram, allowing researchers to stabilize and tune complex broken-symmetry states in the monolayer limit.

One of the phases predicted to emerge from this tuning is "type-II" multiferroicity. Type-II multiferroicity is driven by a complex magnetic texture—typically a spin spiral—that spontaneously breaks spatial inversion symmetry \cite{katsura2005,mostovoy2006,hu2008, tokura_multiferroics_2010}. This mechanism is compelling because it inherently guarantees a strong, intrinsic coupling between the magnetic and electric order parameters. Such magnetoelectric interlocking allows for the manipulation of magnetic states via electric fields, a capability that is essential for the development of energy-efficient spintronic devices \cite{pantel_reversible_2012, hu_progress_2019,gao_two-dimensional_2021}. Theoretical studies confirm that such non-collinear ground states are intrinsic to the TMDH family \cite{Yu2025, riedl2022, sodequist2023, lu2019, Luo2025,Li2023}. However, experimental realization in the monolayer limit remains exceptionally rare, as the balance of magnetic exchange interactions required for this mechanism is easily disrupted by substrate-induced strain, lattice disorder, and charge transfer effects.

Consequently, the first experimental example of intrinsic type-II multiferroicity in the 2D limit was only recently observed in monolayer NiI$_2$, where the state is stabilized by strong iodine-mediated SOC \cite{Ju2021, song2022, amini2024}. Validating whether this magnetoelectric coupling is a robust, tunable feature of the TMDH family—rather than a unique property of the heavy iodide—requires moving beyond this single material example. It is therefore essential to investigate isostructural systems where the relevant energy scales are chemically modified. Monolayer NiBr$_2$ serves as the ideal testbed for this purpose \cite{tokunaga2011, bikaljevic2021, Yu2025, prayitno2024}. The substitution of lighter bromine ligands significantly reduces the spin-orbit coupling and alters the superexchange pathways, effectively shifting the system in the magnetic phase diagram. Investigating NiBr$_2$ thus allows us to test the robustness of the multiferroic ground state against these alterations.

Here, we provide direct experimental evidence that multiferroicity is indeed a general and tunable feature of the nickel dihalide family. Using scanning tunneling microscopy (STM) and spectroscopy (STS), we visualize the ferroelectric domains in monolayer NiBr$_2$ as periodic modulations of the local density of states. We confirm the intrinsic nature of the magnetoelectric coupling through reciprocal control experiments: demonstrating the manipulation of magnetic domains via electric fields and the suppression of electric polarization by external magnetic fields. Furthermore, we map the stability limits of the phase, revealing that the spin-spiral order in NiBr$_2$ is energetically less robust than in its iodide counterpart, consistent with with altered superexchange interactions and the weaker spin-orbit coupling.

\section{Results}

\subsection{NiBr$_2$ Growth and Observation of Multiferroicity}

\begin{figure}[h!]
    \centering
    \includegraphics[width=\textwidth]{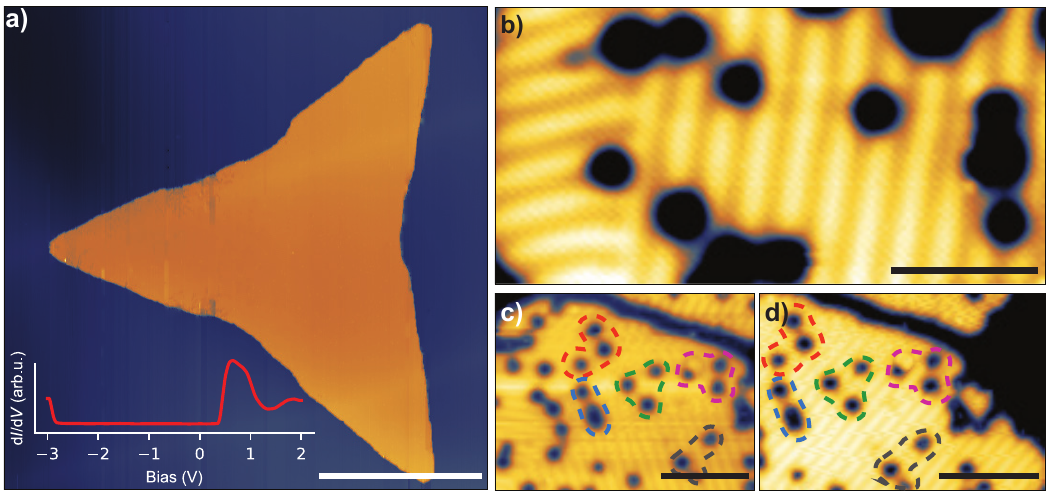}
   \caption{Characterization of monolayer NiBr$_2$. \textbf{a} Overview topography of monolayer NiBr$_2$ island grown on HOPG (\SI{2}{\volt}, \SI{30}{\pico\ampere}, scale bar \SI{100}{\nano\meter}). The inset shows a representative d$I$/d$V$ spectrum taken on a monolayer NiBr$_2$ island. We attribute the small peak at \SI{-2.8}{V} to be the onset of valence band, although are unable to confirm due to junction instabilities around this voltage range. \textbf{b} Zoom-in image of monolayer NiBr$_2$  taken at \SI{0.3}{\kelvin} showing the appearance of stripe-like pattern when scanning within conduction band (\SI{740}{\milli\volt}, \SI{20}{\pico\ampere}, scale bar \SI{10}{\nano\meter}). \textbf{c-d} Electric manipulation of the domains; a NiBr$_2$ island before bias sweep (\SI{740}{\milli\volt}, \SI{20}{\pico\ampere}, scale bar \SI{20}{\nano\meter}) and after bias sweep (\SI{740}{\milli\volt}, \SI{10}{\pico\ampere}, scale bar \SI{20}{\nano\meter}). The circled polarons highlight that the scans have been taken at the same area.} 
    \label{fig:fig1}
\end{figure}

We grow monolayer NiBr$_2$ islands on HOPG by a single-source thermal deposition of NiBr$_2$ on a sample kept at \SI{170}{\degreeCelsius}-\SI{205}{\degreeCelsius} (detailed in the Methods section). By varying the growth parameters, we are able to observe large islands exceeding 200 nm$^2$ as seen in Figure \ref{fig:fig1}a, as well as smaller triangular-shaped islands, which preferentially grow at the substrate step-edges (shown in supplementary information, SI). 

The scanning tunneling spectroscopy (STS) measurements locate the onset of the conduction band (CB) of monolayer NiBr$_2$ on HOPG at around \SI{\sim 0.35}{\volt} (inset of Figure \ref{fig:fig1}a), slightly lower than onset of CB of NiBr$_2$ grown on Au(111) (\SI{\sim 0.8}{\volt}) \cite{bikaljevic2021}. The STS measurements indicate that the valence band (VB) is located around \SI{\sim -2.8}{\volt}, though we were unable to precisely determine the onset due to junction instabilities at large negative voltages. However, theoretical predictions \cite{lu2019} suggest that the band gap of the monolayer NiBr$_2$ is approximately \SI{4}{\volt}, which is consistent with our observation.

Having established the island growth, we proceed to demonstrate the multiferroic nature of monolayer NiBr$_2$. The multiferroicity in monolayer NiBr$_2$ originates from the same mechanism as in monolayer NiI$_2$, which has been investigated both theoretically \cite{fumega2022, riedl2022, sodequist2023, Yu2025, prayitno2024} and experimentally \cite{amini2024,song2022}. In monolayer NiBr$_2$, Ni atoms adopt a 2+ oxidation state and interact via ferromagnetic nearest-neighbor ($J_1$) and antiferromagnetic third-nearest-neighbor ($J_3$) exchange coupling. When the ratio $J_3/J_1$ exceeds a critical value, the competition between exchange interactions favors non-collinear magnetic ordering in the form of a spin spiral characterized by a propagation vector $\mathbf{q}$, whose magnitude depends on $J_3/J_1$ and is given in units of the reciprocal lattice vectors of the structural unit cell with lattice constant $a$.

This spin-spiral ground state, combined with spin-orbit coupling, induces a net electric polarization via the inverse Dzyaloshinskii–Moriya interaction \cite{katsura2005}, making monolayer NiBr$_2$ a type-II multiferroic, where ferroelectricity emerges as a direct consequence of magnetic order. From the relation between magnetization and emergent electric polarization \cite{mostovoy2006,hu2008,fumega2022,amini2024}, it follows that electric polarization is modulated in real space with a periodicity equal to half that of the spin spiral. This, in turn, leads to a modulation of local density of states (LDOS) and the electrostatic potential, which can be detected by STM as a stripe-like contrast with periodicity $\lambda/2$, where $\lambda$ is the magnetic modulation wavelength. STM thus provides a direct means to visualize the multiferroic order without the need for spin-polarized techniques. 

Figure \ref{fig:fig1}b shows the characteristic stripe-like pattern, previously observed for monolayer NiI$_2$ on HOPG \cite{amini2024}, that appears when scanning within the conduction band. As shown in Figure \ref{fig:fig1}b, we regularly observe domain formation with stripes rotated by $60^{\circ}$, corresponding to the threefold symmetry of NiBr$_2$. We identify the charged defects (dark areas in the topography images) as polarons - charge carriers trapped in a lattice distortion that locally bend the conduction band onset upwards (detailed in the SI). \cite{cai2023, Liu2023} Under certain scanning conditions, they can be manipulated across the island, but we have found no evidence of their interplay with the multiferroic order.

As mentioned above, the defining property of type-II multiferroics is the inherent coupling of ferroelectric and magnetic orders, indicating that the magnetic spin-spiral domains are tunable by external electric fields. To demonstrate this coupling in NiBr$_2$, we performed a bias voltage sweep from \SI{1}{\volt} to \SI{10}{\volt} in feedback mode (setpoint \SI{200}{\pico\ampere}) on the HOPG substrate adjacent to the NiBr$_2$ island to mitigate the creation of polarons on the monolayer — a phenomenon also reported for monolayer CoCl$_2$ \cite{cai2023, Liu2023}. Figures \ref{fig:fig1}c and \ref{fig:fig1}d present the NiBr$_2$ monolayer before and after the bias sweep, respectively. The orientation of the multiferroic domains changes distinctively, indicating that the spin-spiral order has been reoriented. This sensitivity of the magnetic structure to electric effects confirms the coupling between electric and magnetic orders, establishing monolayer NiBr$_2$ as a type-II multiferroic.

\subsection{Atomic Scale Characterization}

\begin{figure}
    \centering
    \includegraphics[width=\textwidth]{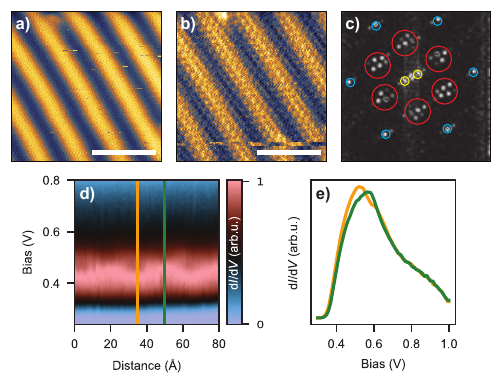}
    \caption{Atomic-scale imaging and spectroscopy of the multiferroic stripes. \textbf{a} High-resolution topography of multiferroic stripes on monolayer NiBr$_2$ acquired within the conduction band (\SI{750}{\milli\volt}, \SI{10}{\pico\ampere}; scale bar: \SI{5}{\nano\meter}). \textbf{b} Topography of the same area taken within the bandgap, where both the atomic lattice and multiferroic stripes are simultaneously resolved (\SI{250}{\milli\volt}, \SI{10}{\pico\ampere}; scale bar: \SI{5}{\nano\meter}). \textbf{c} FFT of image b. Blue circles mark the reciprocal lattice points of the NiBr$_2$ unit cell, yellow circles indicate the multiferroic modulation, and red circles correspond to the moiré pattern arising from the lattice mismatch between HOPG and NiBr$_2$. \textbf{d} Spectroscopic line profile of the conduction band (CB) acquired across the multiferroic stripes. A modulation is visible between \SI{0.5}{\volt} and \SI{0.6}{\volt}. \textbf{e} Individual point spectra taken at the positions indicated by vertical lines in d. A modulation of the LDOS, a shift of the CB maximum, and contrast inversions are observed between \SI{0.4}{\volt} and \SI{0.6}{\volt}.}
    \label{fig:fig2}
\end{figure}

To probe the multiferroic ordering at the atomic limit, we imaged the NiBr$_2$ layer at bias voltages corresponding to the conduction band (\SI{750}{\milli\volt}, Figure \ref{fig:fig2}a) and within the bandgap (\SI{250}{\milli\volt}, Figure \ref{fig:fig2}b). Remarkably, the stripe pattern persists in both energy ranges. Figure \ref{fig:fig2}b, taken within the bandgap, simultaneously resolves the upper bromine atomic lattice and the ferroelectric modulation. This differs from the monolayer NiI$_2$, where multiferroic contrast is only visible when scanning within the conduction band \cite{amini2024}.

From the atomic-resolution topography and its corresponding Fast Fourier Transform (FFT) (Figure \ref{fig:fig2}c), we determine experimental lattice constant of the structural unit cell to be $a = \SI{3.6}{\angstrom}$ and the stripe periodicity to be $L_s = \SI{28}{\angstrom}$. As the observed stripe pattern reflects the modulation of the electric polarization, its periodicity corresponds to half the spin spiral wavelength. This implies a full spin spiral periodicity of \SI{56}{\angstrom}, or approximately 15.1 times the lattice constant ($\lambda = 15.1a$), while the propagation vector $\mathbf{q}$ deviates approximately $7^\circ$ from the $\overline{\Gamma K}$ direction. Notably, we find that the stripe periodicity in NiBr$_2$ is significantly larger than that of NiI$_2$. This increase in $\lambda$ (decrease of $\mathbf{q}$) marks a shift in the ratio of competing exchange constants, placing NiBr$_2$ closer to the transition boundary between the non-collinear spiral and a collinear ferromagnetic state \cite{riedl2022}. The additional spots near the $\Gamma$ point in the FFT (Figure~\ref{fig:fig2}c) are attribuited to moiré pattern arising from the lattice mismatch between the NiBr$_2$ monolayer and the HOPG substrate.

Next, we investigate the local density of states (LDOS) variations associated with the spatial electric modulation. Figure \ref{fig:fig2}d displays scanning tunneling spectroscopy (STS) measurements taken along a line perpendicular to the stripe direction. Unlike monolayer NiI$_2$, where a distinct rigid shift of the conduction band (CB) onset was observed across the stripes \cite{amini2024}, the CB onset in NiBr$_2$ shows negligible variation. Instead, the imaging contrast arises from a shift in the CB maxima and a modulation of the LDOS intensity. Figure \ref{fig:fig2}e presents individual point spectra taken at the extrema of the stripe modulation; while the onsets are nearly identical, the energetic position of the CB maximum shifts by approximately \SI{40}{\milli\volt}. We attribute this LDOS modification to the underlying ferroelectric order, which induces a periodic electrostatic potential that reshapes the local band structure.

\subsection{Phase Stability and Robustness}
\begin{figure}
    \centering
    \includegraphics[width=\textwidth]{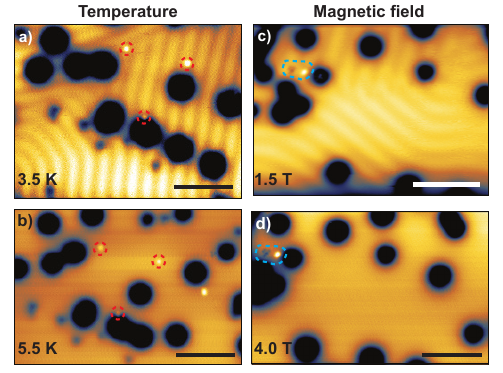}
    \caption{Manipulation of multiferroicity through temperature and magnetic field. \textbf{a-b} Temperature dependence measurement; Scans taken at 3.5 K (\SI{740}{\milli\volt}, \SI{20}{\pico\ampere}, scale bar \SI{10}{\nano\meter}) and at 5.5 K (\SI{740}{\milli\volt}, \SI{20}{\pico\ampere}, scale bar \SI{10}{\nano\meter}). \textbf{c-d} Magnetic field dependence; Scans taken at 1.5 T (\SI{760}{\milli\volt}, \SI{30}{\pico\ampere}, scale bar \SI{10}{\nano\meter}) and at 4.0 T (\SI{760}{\milli\volt}, \SI{30}{\pico\ampere}, scale bar \SI{10}{\nano\meter}). The red (\textbf{a-b}) and blue (\textbf{c-d}) circles show stationery defects, showing that the scans have been taken at same areas.}
    \label{fig:fig3}
\end{figure}

To assess the stability of the multiferroic state and map the phase transition into the ferromagnetic phase, we examined the dependence of the spin-spiral order on temperature and external magnetic fields. The non-collinear magnetic ground state in NiBr$_2$ relies on a delicate balance of exchange interactions and spin-orbit coupling; consequently, thermal fluctuations or the Zeeman interaction are expected to destabilize the spiral propagation \cite{fumega2022, riedl2022, hayami16, tokura_multiferroics_2010, amini2025}. Figure \ref{fig:fig3}a confirms the presence of multiferroic domains at a temperature of \SI{3.5}{\kelvin}. As the temperature is increased, the stripe contrast progressively decreases, disappearing completely at $T_C \sim$ \SI{5.5}{\kelvin} (Figure \ref{fig:fig3}b, Fig.S 2), marking the phase transition from the helimagnetic spin-spiral state to the paramagnetic phase. We subsequently applied an out-of-plane magnetic field to probe the magnetic susceptibility of these domains. While the stripe pattern is unaffected at low fields (Figure \ref{fig:fig3}c), the contrast fades with increasing field strength until the stripes disappear entirely at $B_C \sim$ \SI{4}{\tesla} (Figure \ref{fig:fig3}d, Fig.S 3), indicating that the external field has overcome the frustration of the exchange interactions to force a field-polarized ferromagnetic alignment.

The observed critical values ($T_C \sim$ \SI{5.5}{\kelvin}, $B_C \sim$ \SI{4}{\tesla}) represent a significant reduction in stability compared to monolayer NiI$_2$, where the multiferroic phase persists up to \SI{15}{\kelvin} and remains robust in fields exceeding \SI{11}{\tesla} \cite{amini2025}. This comparative fragility in NiBr$_2$ is attributed to the concerted effect of modified superexchange interactions and the weaker spin-orbit coupling of the lighter bromine ligand \cite{Yu2025, fumega2022, riedl2022, Li2023}. The former sets a lower overall magnetic energy scale bringing the system closer to the ferromagnetic transition boundary (as exemplified by the larger spin spiral periodicity $lambda$), while the latter reduces the anisotropy barrier protecting the non-collinear order against both thermal fluctuations and field-induced alignment. Consequently, whereas high fields in NiI$_2$ only reorient the spiral into a single domain, the less robust magnetic order in NiBr$_2$ allows for a complete suppression of the spin-spiral order at relatively low field strengths.

Finally, these measurements provide further evidence of the magnetoelectric coupling established in the previous sections. Since our STM tip is non-spin-polarized, the contrast in our images arises solely from the electric order — specifically, the modulation of the local electrostatic potential\cite{amini2025}. The fact that an external magnetic field, which couples directly to the atomic spins, suppresses this electric contrast confirms that the electric order is directly linked to the underlying magnetic structure. Together with the electric-field manipulation of the domains shown in Figure \ref{fig:fig1}, this confirms the mutual coupling of the order parameters characteristic of type-II multiferroicity.

\section*{Conclusions}
In conclusion, we have established the type-II multiferroic nature of monolayer NiBr$_2$ on HOPG. STM imaging resolves the ferroelectric domains arising from the non-collinear spin-spiral order, while spectroscopy confirms that this order induces a periodic electrostatic potential modulation that locally alters the LDOS. We demonstrated inherent magnetoelectric coupling through reciprocal manipulation, successfully reorienting magnetic domains via electric fields and suppressing the ferroelectric order with external magnetic fields.

Our observations show that the non-collinear ground state in NiBr$_2$ is shifted in the magnetic phase diagram toward a larger spin-spiral periodicity $\lambda$ and is energetically less robust than in isostructural NiI$_2$. This shift and reduced stability, manifested by lower critical values for $T_C$ and $B_C$, is a direct consequence of the weaker spin-orbit coupling and modified superexchange interactions associated with the lighter bromine ligand. By placing the system closer to the magnetic phase boundary, these effects highlight how the multiferroic energy landscape can be precisely tuned through halide substitution. 

Our findings confirm that the multiferroic mechanism previously observed in NiI$_2$ is a general feature of the transition metal dihalide family. This establishes the dihalides as a versatile platform for chemically engineering magnetoelectric phases at the atomic limit, a critical step toward the realization of programmable low-power spintronic architectures.

\section*{Methods}
\subsection*{Sample preparation} 
The NiBr$_2$ samples were grown in an ultra-high vacuum (UHV) molecular beam epitaxy (MBE) system on highly ordered pyrolytic graphite (HOPG) substrates. The HOPG was exfoliated in ambient conditions and subsequently transferred to the preparation chamber with a base pressure of \SI{2.0e-9}{\milli\bar}, where it was outgassed at \SI{250}{\celsius}. NiBr$_2$ was deposited from a single-source anhydrous powder using a Knudsen cell.

The substrate temperature was \SI{\sim 175}{\celsius} for all samples, with the NiBr$_2$ evaporation temperature \SI{\sim 415}{\celsius}. Growth time varied between 45-50 minutes. Post-annealing was done for 5-15 minutes, around 205-\SI{215}{\celsius}. Our experience indicates that the substrate and evaporation temperature are most critical for reaching the optimal growth. 

\subsection*{STM measurements}
The experiments were carried out with low temperature (LT) UHV Unisoku USM1300 STM system. Base temperature for measurements was \SI{\sim 330}{\milli\kelvin}. All of the measurements were performed with Pt/Ir tips. For the spectroscopy measurements, we use a conventional lock-in technique at a frequency of 757 Hz with \SI{25}{\milli\volt} modulation.  

\section*{Acknowledgements}
We thank Dr. Adolfo Fumega and professor Jose Lado for discussion. 

This research made use of the Aalto Nanomicroscopy Center (Aalto NMC) facilities and was supported by the EU Horizon Europe Marie Skłodowska-Curie Actions (OPTIMISTIC, no.~101109672), European Research Council (ERC-2023-AdG GETREAL, no.~101142364) and the Research Council of Finland (Academy Research Fellow nos.~368478 and 353839, the Finnish Quantum Flagship project no.~358877, and the Finnish Centre of Excellence in Quantum Materials (QMAT)).  

\section*{Author contributions}
Aleš Cahlík: Methodology, Validation, Investigation, Formal analysis, Writing - Original Draft, Supervision, Funding acquisition. Antti Karjasilta: Conceptualization, Methodology, Validation, Investigation, Formal analysis, Writing - Original  Draft. Anshika Mishra: Validation, Investigation. Mohammad Amini: Conceptualization, Investigation. Robert Drost: Formal analysis, Visualization. Javaria Arshad: Investigation. Büşra Arslan: Investigation. Peter Liljeroth: Conceptualization, Supervision, Funding acquisition, Writing - Review and Editing.

\section*{Competing interests}
The authors declare no competing interests.

\phantomsection
\addcontentsline{toc}{section}{\refname}
\bibliographystyle{apsrev4-1} % The native style for RevTeX 4.1
\bibliography{References/reference.bib}

\end{document}